\documentclass{styles/easychair}
\usepackage[utf8]{inputenc}
\usepackage{xcolor}
\usepackage{xspace}
\usepackage{url}
\usepackage{xparse}
\usepackage{pbox,calc}

\usepackage{caption}
\usepackage{subcaption}
\usepackage{algorithm}
\usepackage[ruled,vlined,linesnumbered,algo2e]{algorithm2e}
\usepackage{algpseudocode}

\newcommand{\ourtitle}{Robust Machine Learning for Malware Detection over Time}

\title{\ourtitle}
\date{}

\author{
Daniele Angioni\inst{1}
\and
Luca Demetrio\inst{1,2}
\and
Maura Pintor\inst{1,2}
\and
Battista Biggio\inst{1,2}
}

\institute{
  University of Cagliari,
  Cagliari, Italy\\
  \email{\{daniele.angioni, luca.demetrio93, maura.pintor, battista.biggio\}@unica.it}
\and
   Pluribus One S.r.l.,
   Cagliari, Italy\\
   \email{\{name.surname\}@pluribus-one.it}
 }

\authorrunning{Angioni, Demetrio, Pintor, Biggio}

\titlerunning{\ourtitle}

\begin{document}
\newcommand{\diff}[2]{\frac{\partial #1}{\partial #2}}
\newcommand{\vct}[1]{\ensuremath{\boldsymbol{#1}}}
\newcommand{\mat}[1]{\ensuremath{\mathbf{#1}}}
\newcommand{\set}[1]{\ensuremath{\mathcal{#1}}}
\newcommand{\con}[1]{\ensuremath{\mathsf{#1}}}
\newcommand{\tsum}{\ensuremath{\textstyle \sum}}
\newcommand{\T}{\ensuremath{\top}}

\newcommand{\mycomment}[1]{\textcolor{red}{#1}}
\newcommand{\adcomment}[1]{\textcolor{red}{Ambra: #1}}
\newcommand{\ldcomment}[1]{\textcolor{blue}{Luca: #1}}
\newcommand{\bbcomment}[1]{\textcolor{orange}{Battista: #1}}
\newcommand{\mpcomment}[1]{\textcolor{cyan}{Maura: #1}}
\newcommand{\ascomment}[1]{\textcolor{yellow}{Angelo: #1}}
\newcommand{\frcomment}[1]{\textcolor{yellow}{Fabio: #1}}
\newcommand{\dacomment}[1]{\textcolor{cyan}{Daniele: #1}}

\newcommand{\ind}[1]{\ensuremath{\mathbbm 1_{#1}}}
\newcommand{\argmax}{\operatornamewithlimits{\arg\,\max}}
\newcommand{\erf}{\text{erf}}
\newcommand{\sign}{\text{sign}}
\newcommand{\argmin}{\operatornamewithlimits{\arg\,\min}}
\newcommand{\bmat}[1]{\begin{bmatrix}#1\end{bmatrix}}

\newcommand{\myparagraph}[1]{\smallskip \noindent \textbf{#1}}
\newcommand{\ie}{{i.e.}\xspace}
\newcommand{\eg}{{e.g.}\xspace}
\newcommand{\etal}{{et al.}\xspace}
\newcommand{\etc}{{etc.}\xspace}
\newcommand{\aka}{{a.k.a.}\xspace}
\newcommand{\wrt}{{w.r.t.}\xspace}
\newcommand{\soa}{{state-of-the-art}\xspace}

\newcommand{\ellzero}{$\ell_0$\xspace}
\newcommand{\ellone}{$\ell_1$\xspace}
\newcommand{\elltwo}{$\ell_2$\xspace}
\newcommand{\ellinf}{$\ell_{\infty}$\xspace}

\newcommand{\secsvmcb}{{SVM-CB}\xspace}
\newcommand{\secsvmcbH}{{SVM-CB(H)}\xspace}
\newcommand{\secsvmcbL}{{SVM-CB(L)}\xspace}
\newcommand{\driftanalysis}{{drift-analysis}\xspace}
\newcommand{\tstability}{{T-stability}\xspace}
\newcommand{\nfeatures}{{465,608}\xspace}
\newcommand{\ngoodware}{{116,993}\xspace}
\newcommand{\nmalware}{{12,735}\xspace}
\newcommand{\nandroidapps}{{129,728}\xspace}

\newcommand{\bound}{{r}\xspace}

\newcommand{\IndState}{\State \hspace{\algorithmicindent}}

\newcommand{\maxpages}{10}
\ExplSyntaxOn
\fp_new:N \g_total_sum_fp
\NewDocumentCommand{\AddValue}{m}{
  \fp_gadd:Nn \g_total_sum_fp {#1}
}
\NewDocumentCommand{\DisplaySum}{}{
  \fp_to_decimal:N \g_total_sum_fp
}
\ExplSyntaxOff

\newcommand{\budget}[1]{\textcolor{red}{}}

\maketitle
\begin{abstract}
The presence and persistence of Android malware is an on-going threat that plagues this information era, and machine learning technologies are now extensively used to deploy more effective detectors that can block the majority of these malicious programs.
However, these algorithms have not been developed to pursue the natural evolution of malware, and their performances significantly degrade over time because of such \emph{concept-drift}.

Currently, state-of-the-art techniques only focus on detecting the presence of such drift, or they address it by relying on frequent updates of models.
Hence, there is a lack of knowledge regarding the cause of the concept drift, and ad-hoc solutions that can counter the passing of time are still under-investigated.

In this work, we commence to address these issues as we propose (i) a \driftanalysis framework to identify which characteristics of data are causing the drift, and (ii) \secsvmcb, 
a time-aware classifier that leverages the \driftanalysis information to slow down the performance drop.
We highlight the efficacy of our contribution by comparing its degradation over time with a state-of-the-art classifier, and we show that \secsvmcb better withstands the distribution changes that naturally characterize the malware domain.
We conclude by discussing the limitations of our approach and how our contribution can be taken as a first step towards more time-resistant classifiers that not only tackle, but also understand the concept drift that affects data.
\end{abstract}


\section{Introduction}
\label{intro}
\budget{0.5}
In this information era, we are experiencing tremendous growth in mobile technology, both in its efficacy and pervasiveness.
One of the most common operating systems for mobile devices is Android,~\footnote{\url{https://www.idc.com/promo/smartphone-market-share}} and, because of its popularity, it became particularly attractive to cyber-attackers eyes, who exploit Android vulnerabilities creating malicious applications, also known as \textit{malware}, targeted specifically for these systems~\footnote{https://securelist.com/mobile-malware-evolution-2021/105876/}.
Luckily, the technological development of this era brings enough power to machine learning algorithms, considered the standard for many domains, including cyber-security and, specifically, malware detection, which has shown to be very effective also against never-seen malware families~\cite{arp2014drebin, mariconti2017mamadroid, grosse2017adversarial, ahvanooey2020survey, souri2018state, amamra2012smartphone}.

However, real-world data experience a phenomenon known as \textit{concept drift}, \ie their temporal evolution~\cite{webb2016characterizing}.
In particular, Android applications naturally change over time since attackers keep adjusting malware to bypass detection, and legitimate applications embrace new frameworks and programming patterns while abandoning deprecated technologies.
Recent work highlighted how concept drift worryingly affects the performance of state-of-the-art Android malware detectors, revealing how much it drops over time, contradicting the results achieved by their original analysis since they were inflated by wrong evaluation settings~\cite{pendlebury2019tesseract}.
On top of this issue, the only proposals to counter the concept drift rely on continuous update or retraining of machine learning models~\cite{singh2012tracking, jordaney2017transcend, barbero2020transcending, hu2017concept, narayanan2016adaptive}, instead of tracking which are the characteristics of data that mainly change over time.

Hence, we start bridging the gaps left in the state-of-the-art by proposing novel techniques that understand the concept drift and take advantage of it. 
The contribution of this work are summarized as follows: (i) we propose a \driftanalysis framework that investigates the reasons causing the concept drift inside data, highlighting which 
features are more prone to have a negative contribution to the performance decay;
and (ii) we propose \secsvmcb, a novel classifier that leverages our \driftanalysis information to bound the selected unstable features, reducing the overall performance drop.

We show the effectiveness of \secsvmcb, by comparing its performance over time with Drebin~\cite{arp2014drebin}, a state-of-the-art linear classifier.
To obtain a fair comparison, we train both classifiers on the same dataset, and we show how \secsvmcb better withstand the passing of time, thanks to the domain knowledge acquired through the results of our \driftanalysis framework, thus allowing \secsvmcb to be updated less often compared to Drebin.

We conclude by discussing future directions of this work considering fewer heuristic rules to tune \secsvmcb, and extensions of our methodology to non-linear classifiers.

\section{Android Malware Detection over Time} 
\label{soa}
\budget{1.5}
Before delving into the details of the proposed methods, we firstly describe the structure of Android applications to lay a foundation for understanding the classifier that we consider in this work, and we discuss the problem and proposed solutions to the concept drift problem.

\myparagraph{Android Applications.}
These are programs that run on the Android Operating System. They are distributed as an \textit{Android Application Package} (\texttt{APK}), an archive file with the \texttt{.apk} extension.
An \texttt{APK} contains different files: (i) the \texttt{AndroidManifest.xml}, that stores all the required information needed by the operating system to correctly handle the application at run-time;\footnote{\url{https://developer.android.com/guide/topics/manifest/manifest-intro}} (ii) the \texttt{classes.dex}, that stores the compiled source code and all user-implemented methods and classes; and (iii) additional \texttt{.xml} and resource files that are used to define the application user interface, along with additional functionality or multimedia content.

\begin{figure}
	\centering
	\includegraphics[width=1\columnwidth]{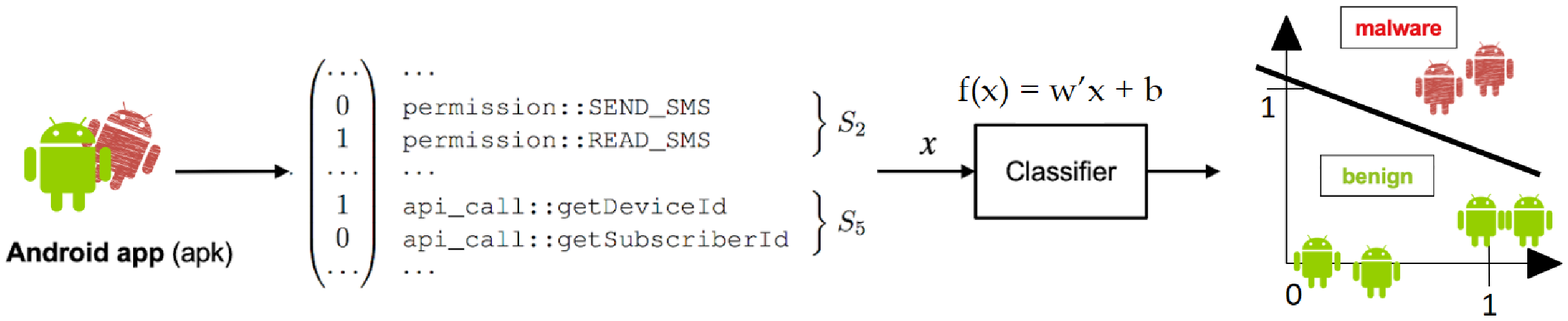}
	\caption{A schematic representation of Drebin~\cite{arp2014drebin}. First, applications are represented as binary vectors in a $d$-dimensional feature space, and this corpus of data is used to train an SVM.
	At test time, unseen applications are fed into a linear classifier, and they are classified as malware if their score $f(x) \geq 0$.}
	\label{img:DREBIN_scheme}
\end{figure}

\myparagraph{Malware Detection with Machine Learning.}
We select a popular binary detector named Drebin~\cite{arp2014drebin} as a baseline for our proposals, for which we show the architecture in Fig .whose architecture is described in Fig. \ref{img:DREBIN_scheme}.
This classifier relies on a Support Vector Machine (SVM)~\cite{cortes1995support} trained on top of hand-crafted features extracted from APKs provided at training time, and they consider: (i) features extracted from \texttt{AndroidManifest.xml}, like hardware components, requested permissions, app components, and filtered intents; (ii) features extracted from \texttt{classess.dex}, including restricted API calls, used permissions, suspicious API calls, and network addresses.
All this knowledge is encoded inside $d$-dimensional feature vectors, whose entries are 0 or 1 depending on the absence or presence of a particular characteristic.
Since Drebin relies on an SVM, it can be used to investigate its decision-making process since each feature is already correlated with a weight that describes its orientation toward one of the two prediction classes, namely legit and malicious.

\myparagraph{Performance over Time.} 
Even though Drebin registered impressive performance in detecting malware, it was not properly tested inside a time-aware environment. I
ts training relies on the Independent and Identical Distribution (I.I.D.) assumption, which takes for granted that both training and testing samples are drawn from the same distribution.
While this property might hold for the image classification domain, it can not be satisfied for the rapidly-growing domain of programs, where training samples differ from future test data as new updates, frameworks, and techniques are introduced while others are deprecated.
The classic evaluation setting injects artifacts inside the learning process, like the presence of samples coming from mixed periods, allowing the classifier to rely on future knowledge at training time.
Such has been demonstrated by Pendlebury et al.~\cite{pendlebury2019tesseract}, that show how selected \soa detectors are characterized by worrying performance drops when evaluated with a more realistic time-aware approach.

\section{Analysing and Improving Robustness to Time}
\label{method}
\budget{2}
We now introduce the two contributions of our work: (i) the \driftanalysis framework to either understand the causes of the concept drift by inspecting the features extracted from data at different time intervals and quantifying their contribution to the overall performance drop; and (ii) the time-aware learning algorithm \secsvmcb (\ie SVM with Custom Bounds), that uses \driftanalysis information to select and bound the weights of a chosen number of features considered unstable to reduce their contribution to the performance decay caused by time.

\myparagraph{Drift-analysis framework.}
Our first contribution tackles the open problem of explaining the concept drift, and we propose the \emph{temporal feature stability} (\tstability), a novel metric measuring the single feature contribution to the performance decay, designed for linear classifiers.
This metric captures two distinct characteristics of each single feature when dealing with time: their relevance in the classifier prediction and their temporal evolution.
These are quantified by the product between (i) the weight $w_j$ corresponding to the $j$-th feature, learned at training time by the classifier; and (ii) the slope $m_j$ that approximates the temporal evolution of the values of the feature.

To compute our metric, we start with the hypothesis that a decrement in the detection rate of malware is strictly related to a decaying score assigned to malware samples as time passes.
Such behavior corresponds to a shift of the malware class distribution towards the decision boundary learned at training time, thus increasing the number of misclassified samples.
To quantify our intuitions, we analyze the variation of the malware score over time, and we compute the conditional expectation of the score over all malware samples (identified with the label $y=1$) at time $t$ as:
\begin{equation}\label{eq:mean score}
	E[\vct w^T\vct x + b|y=1, t] = E\left[ \left( \sum_{j=0}^d w_j \cdot  x_{i,j} \right) + b |y=1, t \right] = \left[ \sum_{j=0}^d w_j \cdot E[x_{i,j}|y=1, t] \right] + b
\end{equation}
where the score is computed as the scalar product between $\vct w$, the vector containing the weights of the linear classifier with bias $b$, and $\vct x_i$ the $d$-dimensional feature vector representation of an input Android application.

Since we want to quantify how the features contribute to the score expectation evolution, we consider the derivative of Eq.~\ref{eq:mean score}, being the summation over the products between weights and the derivatives of the feature expectation \wrt time.
\begin{equation}\label{eq:mean score der}
    \frac{dE[\vct w^T\vct x + b|y, t]}{dt} = \sum_{j=1}^{d-1} w_j \cdot \frac{dE[x_{j}|y, t]}{dt} \approx \sum_{j=0}^{d-1} w_j \cdot m_j = \sum_{j=1}^{d-1} \delta_j
\end{equation}
Since we are interested in capturing the overall trend of the score decay, we approximate each derivative of the $j$-th feature with the slope $m_j$ of the regression line that best fits the single feature expectation over time.
Here, we compress the product $w_j \cdot m_j$ in a single value $\delta_j$, that is how we compute the \tstability of the feature $j$.
Intuitively, the larger and negative the \tstability metric is for a feature, the more such feature accelerates the degradation of the classifier.

Since expectations are not computable for a specific time instant $t$, we quantize the time variable considering time slots with length $\Delta t$, where the $k$-th slot indicate the subset $D_k$ of malware samples registered at time $t \in [k\Delta t, (k+1)\Delta t]$, being $k$ an integer variable.
Thus, we use Alg.~\ref{alg:daf} to obtain the vector $\vct \delta$ containing the \tstability of each feature.
After having computed the number of available time slots $T$ based on the timestamps in $\set{D}$, and the chosen time window $\Delta t$ (line~\ref{line:compute_T}), we initialize a utility matrix $M_{d x T}$ that will contain the mean feature values (line~\ref{line:init_centroids}).
Then we iterate through the time slots (line~\ref{line:iter_k}) and select, for each one, the subset $D_k$ (line~\ref{line:set_Dk}) needed to compute the mean feature value at time $k\Delta t$ storing it in the $k$-th column of $M$ (line~\ref{line:compute_centroid_k}).
After this step, we loop over the number of features (line~\ref{line:iter_j}) to compute the slope $m_j$ of the $j$-th feature over time, \ie the $j$-th row of $M$ (line~\ref{line:compute_slope}), to eventually return the Hadamard product between the classifier trained weights $\vct w$ and the feature slopes $\vct m$ (line~\ref{line:compute_delta}), \ie the \tstability vector $\vct \delta$.

\begin{algorithm}[t]
    \SetKwInOut{Input}{Input}
    \SetKwInOut{Output}{Output}
    \SetKwComment{Comment}{$\triangleright$\ }{}
    \DontPrintSemicolon
    \LinesNumbered
    \Input{The input timestamped and labeled dataset $\set{D}=\{\vct x_i, y_i, t_i\}_{i=1}^n$;
	the time window $\Delta t$;
	the weights $\vct w$ of the reference classifier $g'$.}
    \Output{the T-stability vector $\vct \delta$}
    
    $T \leftarrow \lceil (t_{max}-t_{min})/\Delta t \rceil$ \Comment*[r]{Compute number of time slots}\label{line:compute_T}
    
    $M \leftarrow zeros(d, T)$ \Comment*[r]{Initialize utility matrix}\label{line:init_centroids}
    
    
    \For {$k \in [0, T-1]$}{\label{line:iter_k}
	    $\set{D}_k \leftarrow \{(\vct x_i, y_i, t_i) \in \set{D} : y_i = 1, t_i \in [k\Delta t, (k+1) \Delta t]\}$\label{line:set_Dk}\Comment*[r]{Obtain data in time slot $k$}\label{line:}
	    
	    $M_{*,k} \leftarrow \frac{1}{|D_k|} \sum_{x_i \in D_k} x_{i,*}$\Comment*[r]{Compute mean feature value in time slot $k$}\label{line:compute_centroid_k}
    }
    
    $\vct m \leftarrow zeros(d)$\label{line:init_m}
    
    \For {$j \in [0, d-1]$}{\label{line:iter_j}
        $m_j \leftarrow fit(M_{j,*})$\Comment*[r]{Compute slope of the regression line}\label{line:compute_slope}
    }
    
    $\vct \delta \leftarrow \vct w \circ \vct m$\Comment*[r]{Compute the \tstability vector}\label{line:compute_delta}
    
    \textbf{return} $\vct \delta$ \Comment*[r]{Return \tstability vector}\label{line:return}
\caption{Drift Analysis}
\label{alg:daf}
\end{algorithm}

	
	
	
	    
	    
	    
	    

\myparagraph{Robustness to Future Changes.} 
As our second contribution, we show how to exploit the information obtained with the \driftanalysis inside the optimization process to train \secsvmcb, an SVM classifier hardened against the passing of time.
To train \secsvmcb, we consider a reference temporally unstable classifier to compute the \tstability for each feature.
Then, we select the \textit{unstable features}, that are the $n_f$ of them that have the most negative $\delta_j$ values.
Our goal is to train a new classifier that relies less on these unstable features, thus we bound the absolute value of the correspondent weights to directly reduce their contribution in Eq.~\ref{eq:mean score der}. 
This can be formalized as the constrained optimization problem in Eq.~\ref{eq:objective_function_CB}, where the hinge loss is minimized subject to a constrained on the subset of weights $\set{W}_f$, \ie the $n_f$ weights correspondent to the unstable features, that are forced to be lower than a specific bound $\bound$ in their absolute value.
\begin{eqnarray}\label{eq:objective_function_CB}
\argmin_{\vct{w}, b} && \sum_{i=1}^{n} max(0, 1 - y_i f(\vct{x}_i; \vct{w}, b)),\label{eq:argmin}  \\
s.t. && |w_j| < \bound, \forall w_j \in \set{W}_f.\label{eq:constraint}
\end{eqnarray}

We show in Alg.~\ref{alg:train_daf} the time-aware training algorithm for \secsvmcb that minimize this objective through a gradient descent procedure.
The algorithm is initialized by firstly identifying the subset $\set{W}_f$ of weights corresponding to the $n_f$ unstable features (lines~\ref{line2:init_indexes}-\ref{line2:init_Wf}).
Then, for each iteration, we firstly modulate the learning rate with the function $s(t)$ to improve convergence (line ~\ref{line2:update_eta}), we update the parameters of the classifier to train by applying gradient descent (lines~\ref{line2:update_w}-\ref{line2:update_b}), to eventually clip the weights contained in $\set{W}_f$ to the bound $\bound$ if their absolute value exceed it (line \ref{line2:project}), as described in Eq.~\ref{eq:constraint}.
After $N$ iterations, the algorithm returns the learned parameters $\vct w$ and $b$.
\begin{algorithm}[t]
    \SetKwInOut{Input}{Input}
    \SetKwInOut{Output}{Output}
    \SetKwComment{Comment}{$\triangleright$\ }{}
    \DontPrintSemicolon
    \setcounter{AlgoLine}{0}
    \LinesNumbered
    \Input{$\set{D}=\{\vct{x}_i, \vct{y}_i\}_{i=1}^n$, the training data;
	$\bound$, the absolute value of the bound that must be applied to the weights;
	$\delta$, the T-stability vector; 
	$n_f$, the number of weights  that must be bounded;
	$N$, the number of iterations;
	$\eta^{(0)}$, the initial gradient step size ;
	$s(t)$ a decaying function of $t$.}
    \Output{$\vct{w},b$, the trained classifier's parameters.}
    
    $\set{J} \leftarrow argsort(\vct \delta)$\Comment*[r]{Initialize feature indexes ordered \wrt $\delta$}\label{line2:init_indexes}
    
	$\set{J}_f \leftarrow \{j_k:  k=0,...,n_f\}, j_k \in \set{J}.$\Comment*[r]{Select first $n_f$ indexes}\label{line2:nf_indexes}
	Initialize $\set{W}_f = \{w_j: j \in  \set{J}_f\}$\Comment*[r]{Select corresponding $n_f$ weights}\label{line2:init_Wf}
	

	$(\vct{w}^{(0)}, b^{(0)}) \leftarrow (\vct0, 0)$\Comment*[r]{Initialize parameters}\label{line2:init_params}

	\For {$t \in [1, N]$}{\label{line2:iterations}
		$\eta^{(t)} \leftarrow \eta^{(0)} s(t)$\Comment*[r]{Update learning rate}\label{line2:update_eta}
		
		$\vct{w}^{(t)} \leftarrow \vct{w}^{(t-1)} - \eta^{(t)} \nabla_{\vct{w}} \set{L}$\Comment*[r]{Update weights}\label{line2:update_w}
		
		$b^{(t)} \leftarrow b^{(t-1)} - \eta^{(t)} \nabla_{b} \set{L}$\Comment*[r]{Update bias}\label{line2:update_b}
		

        
        $\vct{w}^{(t)} \leftarrow Clip(\vct{w}^{(t)}; \set{W}_f, \bound)$ \Comment*[r]{Clip weights based on Eq.~\ref{eq:constraint} criteria}\label{line2:project}
	}
	
	

    \textbf{return} $\vct{w}^{(t)}, b^{(t)}$ \Comment*[r]{Return the learned parameters}\label{line2:return}
    
\caption{\secsvmcb learning algorithm}
\label{alg:train_daf}
\end{algorithm}

\section{Experiments}
\label{exp}
\budget{4}
We now apply our methodology to quantify how it explains and hardens a classifier against the performance decay compared with the time-agnostic classifier Drebin~\cite{arp2014drebin}.

\begin{figure}
	\centering
	\includegraphics[width=1\columnwidth]{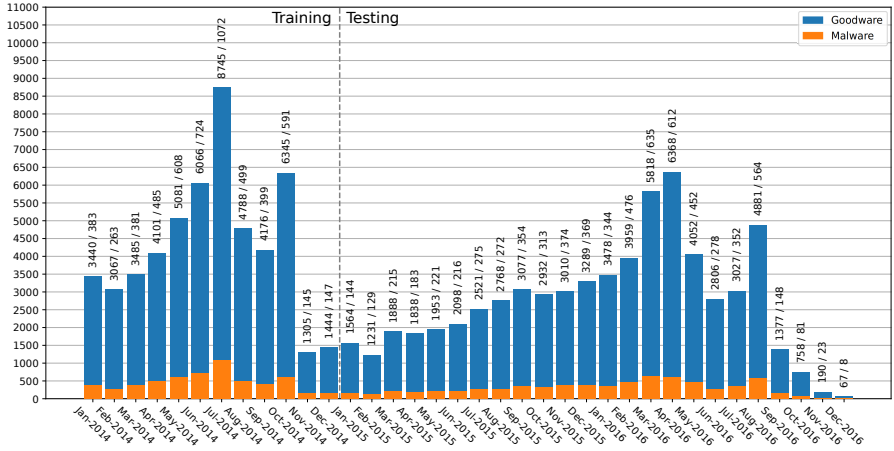}
	\caption{Stack histogram with the monthly distribution of apps, spanning from Jan 2014 to Dec 2016. The dashed vertical lines determine the considered time-aware temporal split.}
	\label{img:priors}
\end{figure}

\myparagraph{Dataset.} 
We leverage the dataset provided by Pendlebury et al.~\cite{pendlebury2019tesseract}, composed of \ngoodware legitimate and \nmalware malicious Android applications sampled from the AndroZoo dataset~\cite{allix2016androzoo}, spanning from January 2014 to December 2016.
We replicate their temporal train-test split as shown in Fig. \ref{img:priors}, by dividing them between December 2014 and January 2015, and we set the time slot $\Delta t$ equal to $1$ month to ensure sufficient statistics for each.
We hence extract \nfeatures from the training set to match the original formulation of Drebin~\cite{arp2014drebin}.

\myparagraph{Models.}
We consider Drebin as the baseline classifier, trained with the $C$ parameter set to $1$, and we compare it with two versions of \secsvmcb by considering different bounds on the unstable features detected by the \driftanalysis framework. 
We will refer to the baseline classifier as SVM since the underlying feature extractor and the feature embedding module are the same for all the classifiers under analysis.

\begin{figure}
	\centering
	\begin{subfigure}{0.32\textwidth}
        \includegraphics[width=1\linewidth]{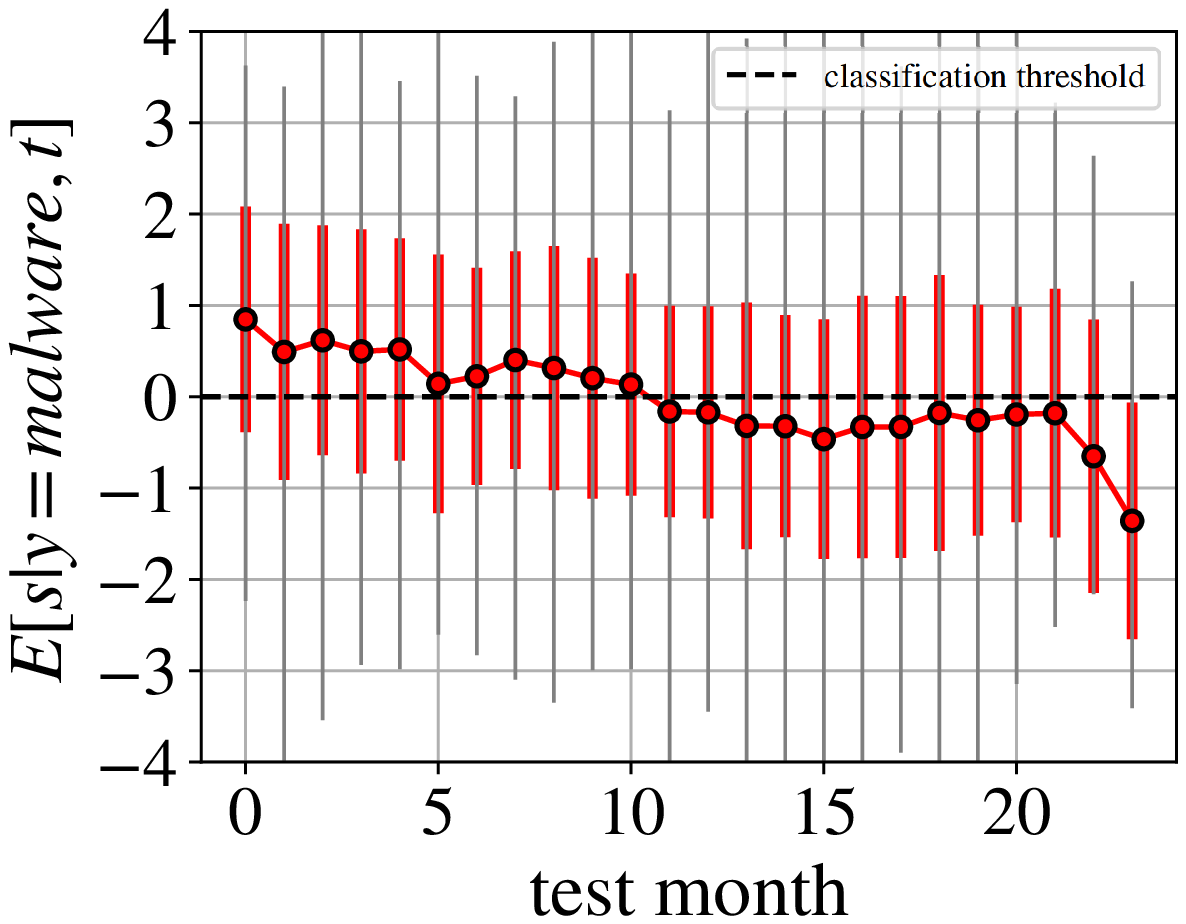}
        \caption{}
        \label{img:score_malware}
    \end{subfigure}
    \begin{subfigure}{0.32\textwidth}
        \includegraphics[width=1\linewidth]{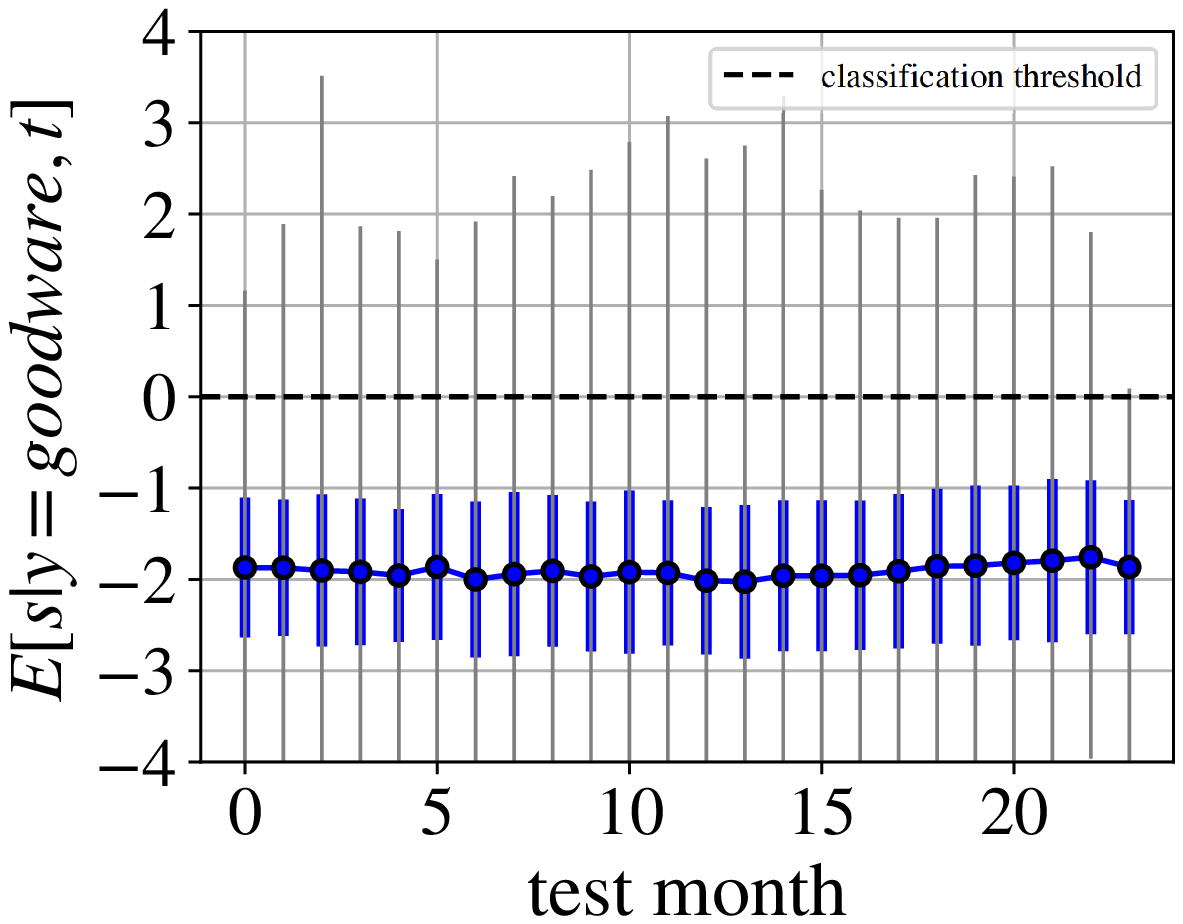}
        \caption{}
        \label{img:score_goodware}
    \end{subfigure}
    \begin{subfigure}{0.32\textwidth}
        \includegraphics[width=1\linewidth]{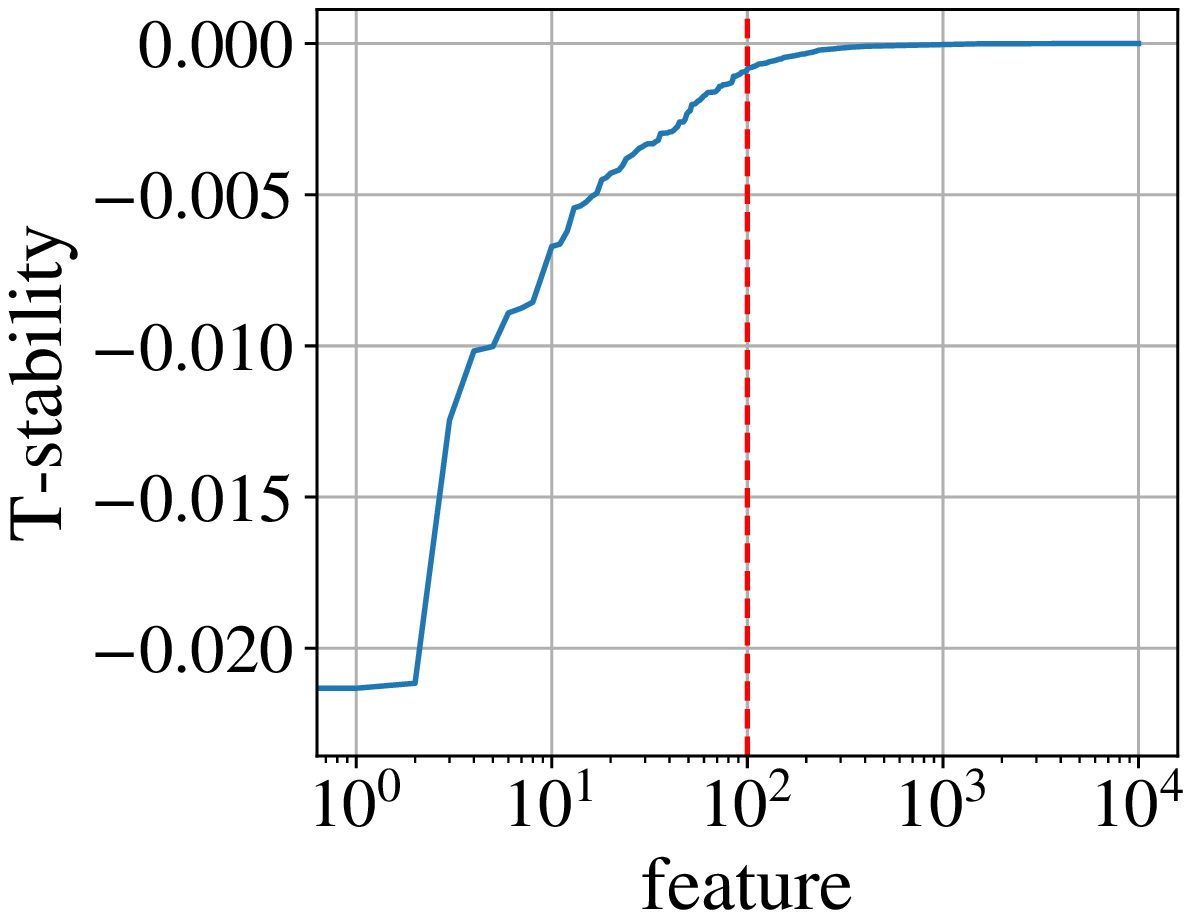}
        \caption{}
        \label{img:deltas}
    \end{subfigure}
	\caption{The mean score over the testing periods assigned by the SVM to malware (a) and goodware samples (b) of the test set, along with their standard deviation (colored thick lines) and min-max range(thin grey lines).
	Lastly, the $10^4$ \tstability values in increasing order, computed through Alg.~\ref{alg:daf} (c).}
	\label{img:score_vs_time_SVM}
\end{figure}

\myparagraph{Drift Analysis Results.}
To identify the features responsible for the performance decay over time in our baseline SVM, we firstly show in Fig.~\ref{img:score_vs_time_SVM} the trend of the mean score assigned respectively to malicious (Fig.~\ref{img:score_malware}) and benign samples (Fig.~\ref{img:score_goodware}) over all the testing periods.
While the classifier assigns, on average, an almost constant negative score to the goodware class, the mean score assigned to malware gradually approaches to zero to eventually become negative after 10 months, thus validating the hypothesis claimed in Sect.~\ref{method}.

We compute the \tstability vector $\vct \delta$ through Alg. \ref{alg:daf} for the learned weights of the SVM \wrt the timestamped training set, and we show the first $10^4$ \tstability values in increasing order along with the corresponding features in Fig.~\ref{img:deltas}.
The latter highlights that most of the contribution to the performance decay is caused by roughly 100 features among all the feature set, while all the remaining ones do not substantially compromise the detection rate over time since their \tstability is very close to zero.

We report a subset of the selected unstable features (\ie features presenting large negative \tstability values) in Table~\ref{tab:featurenames}.
The first 10 rows show features that the SVM associates with the goodware class and are becoming more likely to be found in malware ($w_j < 0, m_j > 0$), while the last 10 rows show features that the SVM associates with the malware class but they are disappearing from data ($w_j > 0$, $m_j < 0$).
For simplicity, we will refer to the features in the first and second table, respectively, as the first and the second group of features.

We can recognize in the first group features mostly related to commonly-used URLs.
For instance, among them, we find ``www.google.com'', ``www.youtube.com'', and websites under the ``facebook.com'' domain, which are all legitimate URLs to browse, and the classifier links them to the goodware class by assigning them a positive weight.
The second group is mostly characterized by features related to intents and activities. 
For instance, we find the presence of a cipher algorithm (``interesting\_calls::Cipher(DES)''), reported to be used to obfuscate and encrypt part of the malicious application.\footnote{\url{https://www.virusbulletin.com/virusbulletin/2014/07/obfuscation-android-malware-and-how-fight-back}}
However, this feature has a decreasing trend ($m_j < 0$), meaning that malware relies less on this method as time passes, probably because it would ease the detection of the malware under manual inspection.

From this analysis, we can deduce that the unstable features can be grouped into two types of features: (i) goodware-related features that malware creators are starting to inject in their malicious code to increase the probability of it being recognized as goodware, and (ii) malware-related features that malware creators are starting to deprecate to reduce the probability of it being recognized as malware.

\begin{table}[]
    \caption{List of 20 features taken from the set of unstable features. 
    The first column contains the considered features, the second column represents their \tstability measure $\delta_j$, the third column the weight $w_j$ assigned by the SVM, and the fourth column is the estimated angular coefficient $m_j$.
    The first 10 rows show goodware-related features which are becoming more frequent in malware as time passes, while the last 10 rows show malware-related features which are disappearing from this class.}
    \label{tab:featurenames}
    \centering
    \resizebox{\textwidth}{!}{
    \begin{tabular}{l c c c }
    \hline \hline
    Feature name &  $\delta_j$ &         $w_j$ &         $m_j$ \\
    \hline \hline
    urls::https://graph.facebook.com/\%1\$s?...\&accessToken=\%2\$s & -0.008753 & -0.596730 &  0.014669 \\
    intents::android\_intent\_action\_VIEW               & -0.010168 & -0.462059 &  0.022005 \\
    urls::http://www.google.com 	& -0.021320 & -0.436577 &  0.048835 \\
    activities::com\_revmob\_ads\_fullscreen\_FullscreenActivity & -0.006204 & -0.348884 &  0.017782 \\
    activities::com\_feiwo\_view\_IA                      & -0.004435 & -0.347665 &  0.012758 \\
    urls::http://i.ytimg.com/vi/                       & -0.005245 & -0.319063 &  0.016438 \\
    api\_calls::android/content/ContentResolver;$\rightarrow$openInputStream  & -0.003749 & -0.302131 &  0.012410 \\
    urls::https://m.facebook.com/dialog/              & -0.004955 & -0.285100 &  0.017379 \\
    urls::http://market.android.com/details?id=        & -0.004041 & -0.260522 &  0.015510 \\
    urls::http://www.youtube.com/embed/                & -0.004289 & -0.259927 &  0.016502 \\
    \hline
    api\_calls::android/net/wifi/WifiManager;$\rightarrow$getConnectionInfo & -0.003469 &  0.148022 & -0.023438 \\
    app\_permissions::name='android\_permission\_MOUNT\_UNMOUNT\_FILESYSTEMS' & -0.004508 &  0.296193 & -0.015220 \\
    urls::http://e.admob.com/clk?... & -0.006713 &  0.427714 & -0.015695 \\
    activities::com\_feiwothree\_coverscreen\_SA          & -0.003564 &  0.443662 & -0.008034 \\
    interesting\_calls::Cipher(DES)                     & -0.008910 &  0.489497 & -0.018202 \\
    intents::android\_intent\_action\_PACKAGE\_ADDED       & -0.022435 &  0.702801 & -0.031922 \\
    activities::com\_fivefeiwo\_coverscreen\_SA           & -0.003813 &  0.743198 & -0.005131 \\
    intents::android\_intent\_action\_CREATE\_SHORTCUT     & -0.012456 &  0.748091 & -0.016650 \\
    intents::android\_intent\_action\_USER\_PRESENT        & -0.021155 &  0.803000 & -0.026344 \\
    activities::com\_feiwoone\_coverscreen\_SA            & -0.010022 &  1.141652 & -0.008778 \\
    \hline
    \end{tabular}
    }
\end{table}

\begin{figure}
	\centering
    \begin{subfigure}{0.24\textwidth}
        \includegraphics[width=1\linewidth]{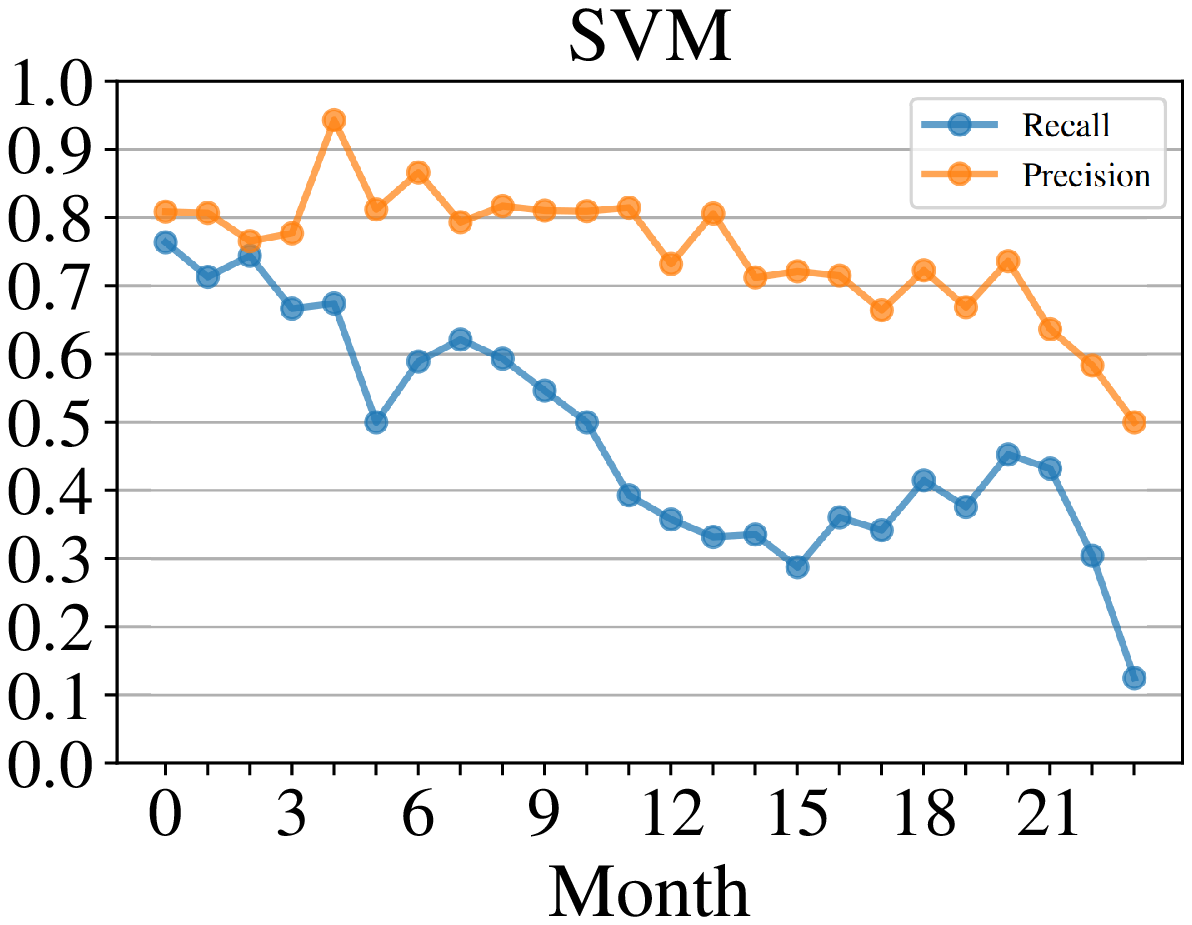}
        \caption{}
        \label{img:decay_svm}
    \end{subfigure}
	\begin{subfigure}{0.24\textwidth}
        \includegraphics[width=1\linewidth]{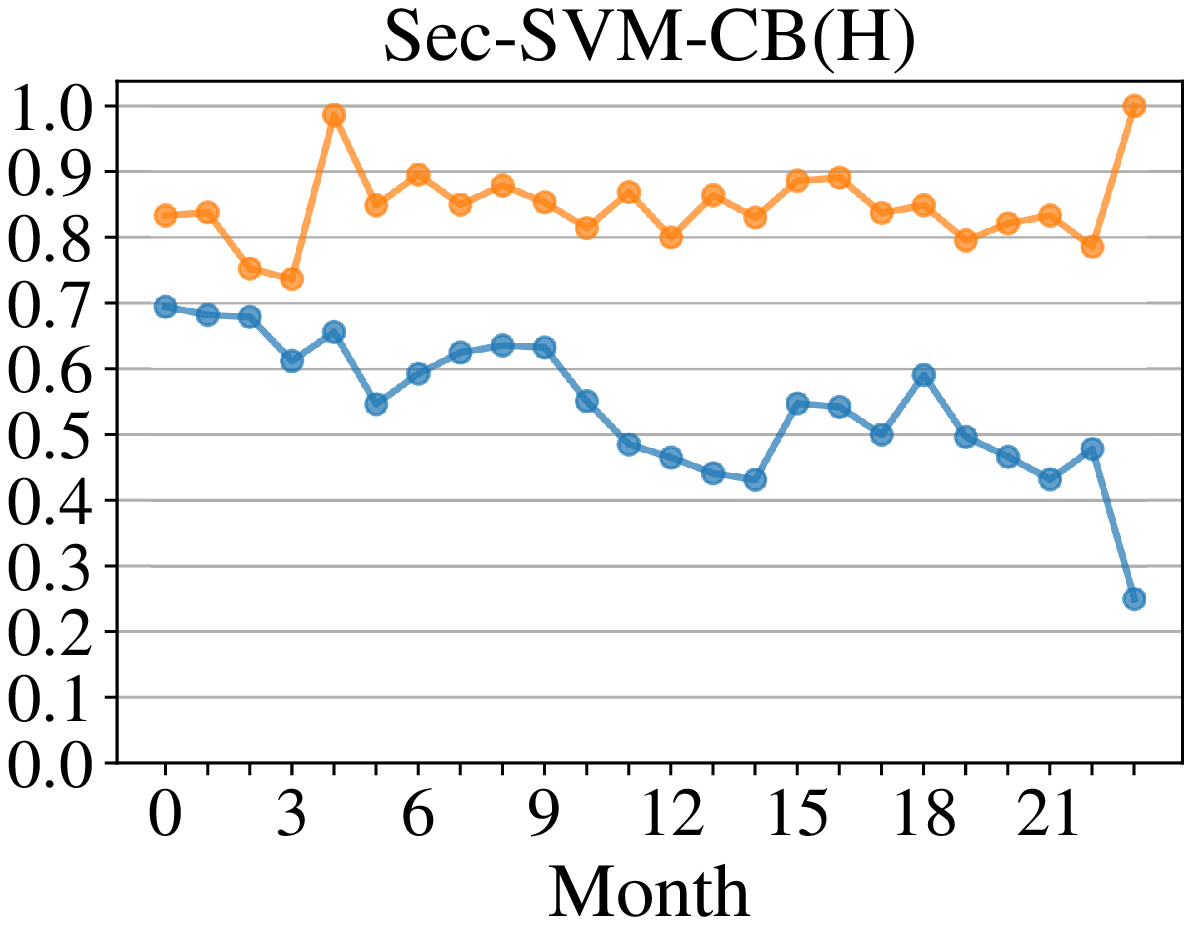}
        \caption{}
        \label{img:decay_svmh}
    \end{subfigure}
    \begin{subfigure}{0.24\textwidth}
        \includegraphics[width=1\linewidth]{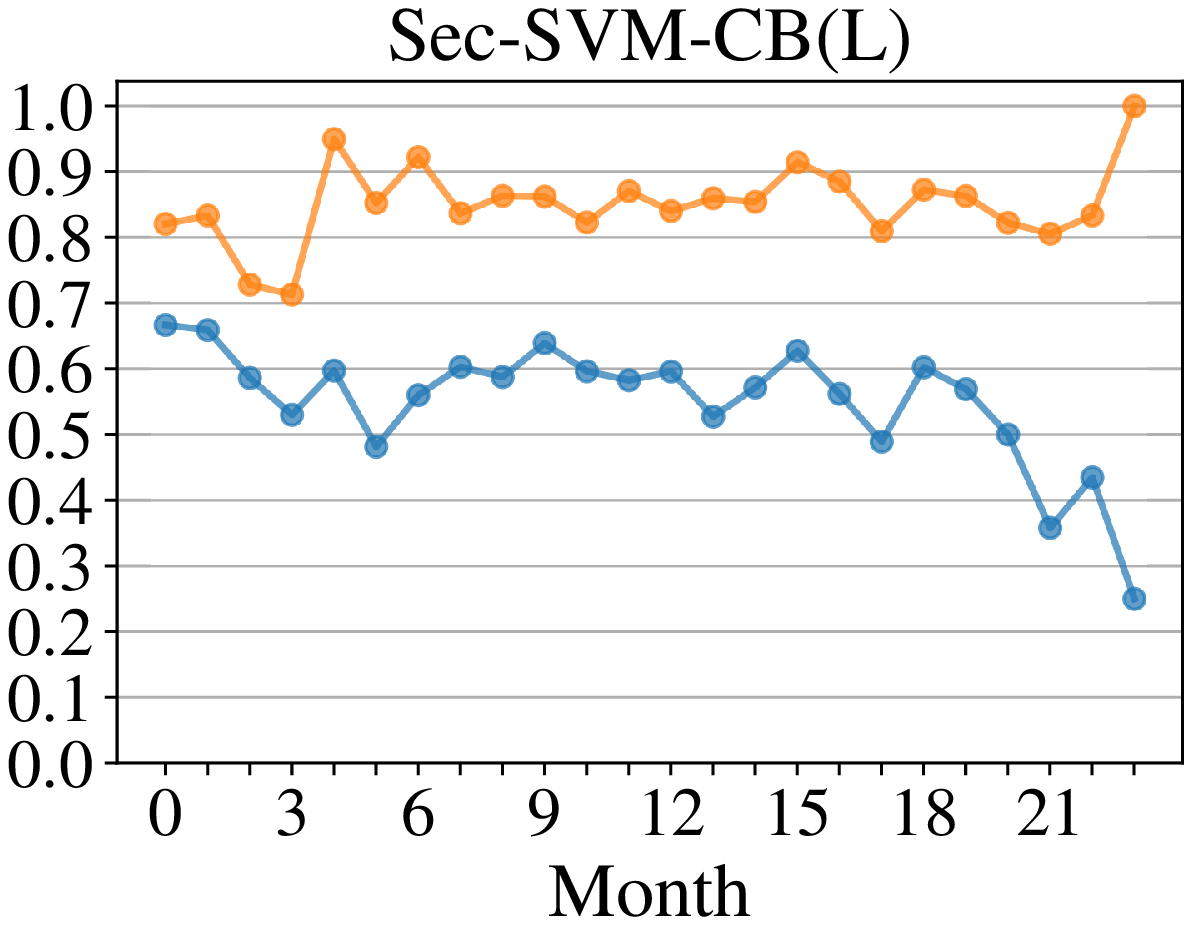}
        \caption{}
        \label{img:decay_svml}
    \end{subfigure}
    \begin{subfigure}{0.24\textwidth}
        \includegraphics[width=1\linewidth]{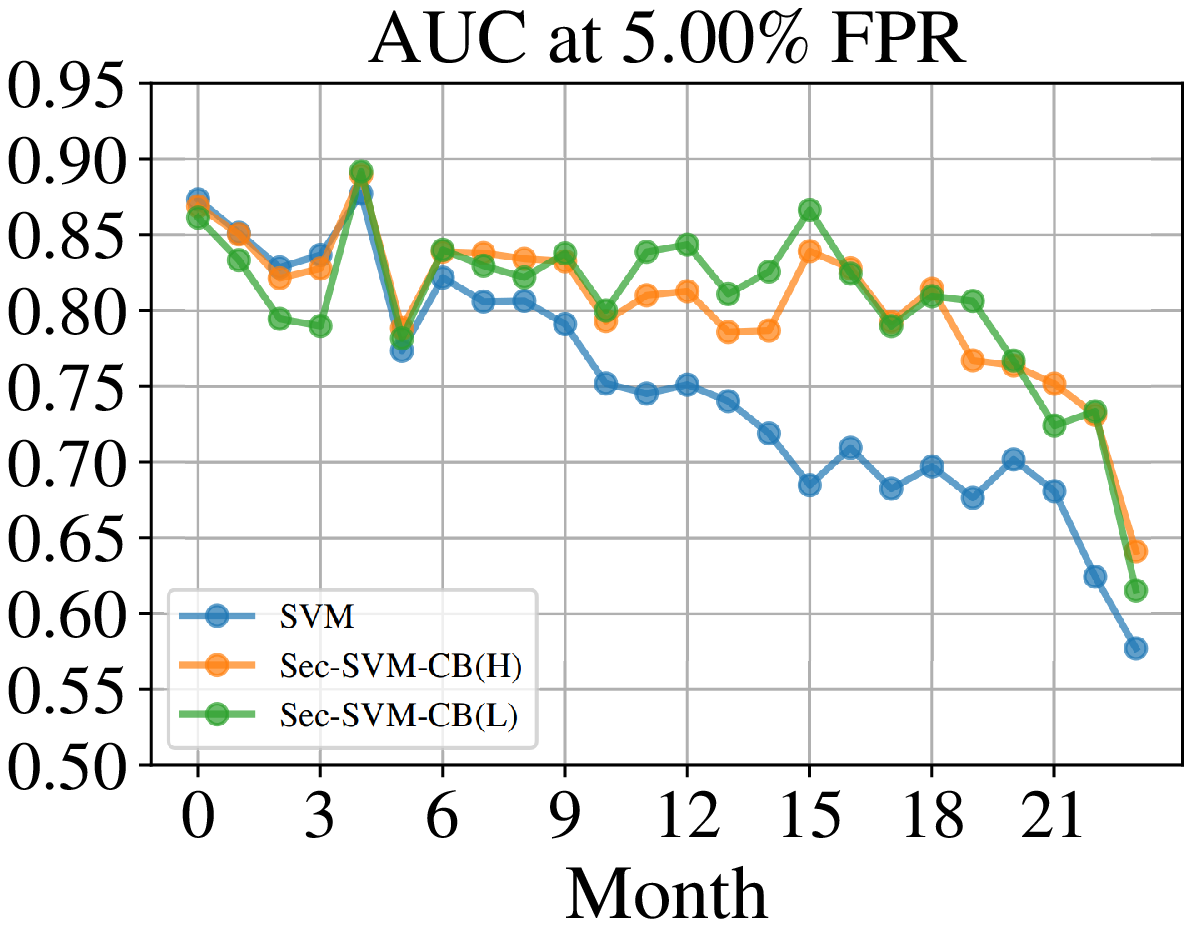}
        \caption{}
        \label{img:decay_auc}
    \end{subfigure}
	\caption{The precision (orange) and recall (blue) of SVM (a), \secsvmcb(H) (b) and \secsvmcb(L) (c), and the Area Under the ROC curves (AUC) at 5\% (d) for the three classifiers over the 2-years testing periods (from Jan-2015 to Dec-2016).}
	\label{img:perf_decay_auc}
\end{figure}

\myparagraph{Improving Robustness.}
We now leverage the results of our \driftanalysis framework performed on the SVM by training \secsvmcb using Alg.~\ref{alg:train_daf}, running it for 2000 iterations, with the initial learning rate $\eta^{(0)}$ set to $7 \cdot 10^{-5}$ and we use the cosine annealing function as $s(t)$ to modulate it over the iterations.
We heuristically choose the number of features to bound $n_f = 10^2$, since these are the ones the most contribute to the performance decay (Fig.~\ref{img:deltas}).
We train two versions of \secsvmcb, referred as (i) \secsvmcbH the classifier with $r=0.8$ and (ii) \secsvmcbL the classifier with $r=0.2$.
These two different bounds allow us to better understand how the robustness against the concept drift changes when we apply softer ($r=0.8$) or harder ($r=0.2$) constraints to the correspondent weights.
We report the performance analysis of these classifiers in Fig. \ref{img:perf_decay_auc}, where we show the evolution over the testing periods of the recall (red) and the precision (blue) for the SVM (Fig.~\ref{img:decay_svm}) and \secsvmcb(L-H) (Fig.~\ref{img:decay_svml} and~\ref{img:decay_svmh}). 
We will focus mainly on the discussion of the recall curves, as our primary concern is the detection rate of the malware samples over time, which is computed in the same way.
Also, we will not discuss the results concerning the last two months, as the number of samples is not sufficiently large for a proper evaluation (as highlighted by Fig.~\ref{img:priors}).

We correctly replicated the results obtained by Pendlebury et al.~\cite{pendlebury2019tesseract} for the SVM, which presents the highest recall among the tested classifiers in the first testing periods, starting from 76.4\%, dropping fast towards a 28.8\% recall at 16-th month to eventually rise to 45.3\% at 21-th month.
Although the initial detection rate of \secsvmcbL is lower than 70\% it fluctuates less \wrt to the baseline by maintaining the performance around 50-60\% with a final drop to 35.8\% at the third to last month.
\secsvmcbH presents an initial recall of 69.4\%, while it decays to 43.2\% once it reaches the 22-th month.
Coherently to the results obtained by Pendlebury et al.~\cite{pendlebury2019tesseract}, we observe that the baseline SVM is characterized by the fastest performance decay, while the other classifiers
start between 60\% and 70\% recall.
The peak of temporal robustness is reached by \secsvmcbL where the recall curve seems to be almost flattened, while \secsvmcbH has indeed a slower decay \wrt the SVM but faster than \secsvmcbL.
Lastly, Fig. \ref{img:decay_auc} shows the Area Under the Receiving Operating Curve (ROC) curve for each testing period, computed up to $5\%$ FPR.
Here we indirectly discuss the correlation between precision and recall considering the performance when we fix a constant percentage of goodware misclassified as malware for each month in order to better measure and compare the data separation capabilities of the three classifiers. 
The AUC curves reflect what we have discussed for the recall: the SVM starts as usual with the highest AUC and decays rapidly below all the other AUC curves, while the other classifiers start with a lower AUC that reveals to be higher than SVM when approaching the 10-th month.
\secsvmcbL has been confirmed to be the more stable classifier even in this constrained evaluation setting with low FPR. 

\section{Related Work}
\label{rel}
\budget{1}
We now offer an overview of state-of-the-art techniques similar to our proposal.
Pendlebury et al.~\cite{pendlebury2019tesseract} proposes Tesseract, a test-time evaluation framework to determine the faultiness of classifiers in the presence of the concept-drift.
The authors show that evaluations are affected by misleading biases that inject artifacts inside the trained machine learning model, thus causing a performance decay once the model faces real-world data.
Tesseract highlights how different proposed models do not cope with the concept drift of Android applications and that faulty training settings inflated their original evaluations.
While Tesseract is a consistent method to include concept drift in the evaluation, it is not designed to either fix or mitigate its presence.

Jordaney et al.~\cite{jordaney2017transcend}, propose Transcend, a framework that signals the premature aging of classifiers before their performance starts to degrade consistently by analyzing the difference between samples observed at training at test time.
On top of this methodology, Barbero et al.~\cite{barbero2020transcending} propose Transcendent, which improves Transcend to include the rejection of out-of-distribution samples that cause the performance drops.
However, they do not propose methods to harden a classifier against concept drift, rather they focus on protection systems exploiting samples encountered during deployment, such as a notification when data start differing from the training one~\cite{jordaney2017transcend}, or directly rejecting a sample coming from a drifted data distribution~\cite{barbero2020transcending}.

In contrast to previous work, we consider the presence of faulty evaluations, and we extend it with a methodology that quantifies which features of the data distributions are changing and how.
Such contribution not only explains the performance decay, but also helps understanding the reasons behind the concept drift.
Instead of rejecting samples or just signaling the worsening of the performances of a model, we build a time-aware classifier that takes into account the acquired knowledge of the data distribution changes, and we show how our methodology can better withstand the passing of time.

\section{Conclusions and Future Work}
\label{concl}
\budget{0.5}
In this work, we propose a preliminary methodology that understands and provide an initial hardening against the concept drift that plagues the performance of Android malware detection.
In particular, we develop a \driftanalysis framework that highlights which features contribute more to the performance decay of a classifier over time, and we leverage these results to propose \secsvmcb, a linear classifier hardened against the passing of time.

We show the efficacy of our proposals by applying our drift-analysis framework to Drebin, a linear Android malware detector, and we compare its performances over time against its hardened version computed through our proposed methodology.
From our experimental analysis, we can precisely detect which features worsen the detection rate of Drebin and how the trained \secsvmcb better withstand the passing of time.
In particular, we highlight the efficacy of the bounding of these unstable features, reducing the performance drop of \secsvmcb w.r.t. the baseline Drebin.

Although the obtained results are promising, this work presents the following limitations.
First, the experimental setup does not guarantee that the provided solution against performance decay can be applied to other types of detectors, as this work addresses the problem of analyzing the effect of the concept drift only for linear classifiers that work only on static features~\cite{arp2014drebin, demontis2017yes}.
Also, the \tstability might not reflect the actual concept drift that affects Android applications, as it is computed on a classifier trained on a specific dataset, which approximates the real data distribution.
Hence, we should also study the Android malware domain more to provide sufficient and reliable evidence of why the features chosen by the drift-analysis framework are actually causing the decay.
Lastly, we heuristically tuned the bounds for the selected weights of \secsvmcb, but these choices could be improved with an automatic algorithm that computes the ones that lead to better robustness against time.

However, we anyhow believe that our work can suggest a promising research direction that will provide more insight on the usage of each contribution.
We first intend to explore more advanced methods based on the drift-analysis framework, including an automatic bound selection for the weights inside the learning algorithm, by adopting a regularization term tailored specifically for temporal performance stability.
Secondly, we intend to generalize this method to address deep learning algorithms, where the feature extractor and the feature representation of the last linear layer evolve during training.

Moreover, we will explore other research directions, such as (i) the quantification and prevention of machine learning malware detectors from forgetting old threats when updated with new data, 
and (ii) the inclusion of research fields such as Continual Learning,~\footnote{https://www.continualai.org/} which model data as a continuous stream, thus enabling the development of techniques for updating classifiers constantly and effortlessly.

\section*{Acknowledgments}
This work has been partly supported by the PRIN 2017 project RexLearn, funded by the Italian Ministry of Education, University and Research (grant no. 2017TWNMH2); and by the project TESTABLE (grant no. 101019206), under the EU’s H2020 research and innovation programme.

\bibliographystyle{unsrt}
\bibliography{bibliography}

\end{document}